\documentclass[9pt,twocolumn,twoside]{wlscirep}
\usepackage{float}
\title{Unraveling nonadiabatic ionization and Coulomb potential effect in strong-field photoelectron holography}

\author[1]{Xiaohong Song}
\author[1]{Cheng Lin}
\author[1]{Zhihao Sheng}
\author[1]{Peng Liu}
\author[1]{Zhangjin Chen}
\author[1,*]{Weifeng Yang}
\author[2,3]{Shilin Hu}
\author[4]{C. D. Lin}
\author[2,3,*]{Jing Chen}

\affil[1]{Department of Physics, College of Science, Shantou University, Shantou, Guangdong 515063, People's Republic of China}
\affil[2]{HEDPS, Center for Applied Physics and Technology, Peking University, Beijing 100084, People's Republic of China}
\affil[3]{Institute of Applied Physics and Computational Mathematics, P. O. Box 8009, Beijing 100088, People's Republic of China}
\affil[4]{J.R.Macdonald Laboratory, Physics Department, Kansas State University, Manhattan,Kansas 66506-2604, USA}
\affil[*]{Corresponding author: wfyang@stu.edu.cn, chen\_jing@iapcm.ac.cn}

\begin{abstract}
	 Strong field photoelectron holography has been proposed as a means for interrogating the spatial and temporal information of electrons and ions in a dynamic system.  After ionization, part of the electron wave packet may directly go to the detector (the reference wave), while another part may be driven back to the ion where it  scatters off (the signal wave). The interference hologram of the two waves may be used to retrieve the target information. However, unlike conventional optical holography, the propagations of electron wave packets are affected by the Coulomb potential as well as by the laser field. In addition,  electrons are emitted over the whole laser pulse duration, thus multiple interferences may occur. In this work, we used a generalized quantum-trajectory Monte Carlo method to investigate the effect of Coulomb potential and the nonadiabatic subcycle ionization on the photoelectron hologram.  We showed that photoelectron hologram can be well described only when the nonadiabatic effect in ionization is accounted for, and Coulomb potential can be neglected only in the tunnel ionization regime. Our results help establishing photoelectron holography for probing spatial and dynamic properties of atoms and molecules.
\end{abstract}
\begin{document}

\maketitle
\thispagestyle{fancy}

\section{Introduction}

Atomic photoionization under intense laser irradiation is a
fundamental process in strong-field light-matter interaction.
Since above-threshold ionization (ATI) was firstly observed more
than thirty years ago \cite{ap79}, a series of experimental
discoveries together with subsequent theoretical efforts have
greatly advanced our understanding of the underlying physics of
laser-atom interactions
\cite{wbecker1,Corkum1}. In recent years,  with the availability of new long wavelength lasers and high resolution electron spectrometers, photoelectron spectra from some recent experiments have revealed a number of surprises. Besides the familiar ATI peaks,  many new additional ``peaks or fringes" have been observed in the two-dimensional electron momentum spectra.  These new features, usually called by some new acronym or simply by ``structures", appear to be quite general, as they are nearly independent of the target atoms or molecules, yet they are dependent of the laser wavelength, intensity and sometimes also of pulse duration.  Among these new discoveries are the so-called ``low-energy structures" (LES) at a few eV or sub-eV's or the ``very low-energy structure" (VLES) at a few meV, above the ionization threshold\cite{Blaga2009NatPhys,Quan2009PRL,Wu2012PRL,liu10,yan10,rost12,chu2012PRA,Guo2013}. The widely familiar strong field approximation (SFA) in most cases are incapable of interpreting these observations. It is intuitively clear that a quantitative theory for such low energy electrons would require the incorporation of Coulomb potential from the ion core.   On the other hand, there are other higher energy features \cite{Huismans7,Hickstein8,Meckel9,Huismans10,Marchenko11,Yang12,Yang14,Bian2011} which lie close to the so-called 2$U_p$ cutoff  limit  ($U_p$ is the ponderomotive energy or the averaged quiver energy of a free electron in the laser field). Among them we will focus on the so-called ``side lobes" observed in the photoelectron momentum distribution
(PMD). Such side lobes were observed in the PMD of metastable xenon atoms ionized with intense 7000 nm laser pulses from a free-electron laser \cite{Huismans7}. They have been further observed or found in numerical calculations for other wavelengths. These side lobes were interpreted as analogous to optical holograms, resulting from the interference of the direct ionized electron wave packet (reference wave) and the laser-driven scattered electron wave packet (the signal wave) where the electron has been scattered off the target ion once before reaching the detector. Like holography, such interference may encode target structure information.

It is well known that atomic photoionization in intense laser
field can be categorized into two regimes: tunneling regime and
multiphoton regime according to the Keldysh parameter
$\gamma=\sqrt{\frac{I_{p}}{2U_{p}}}$ ($I_{p}$ is the ionization
potential,  $U_{p}=\frac{I}{4\omega^{2}}$ is the
ponderomotive energy,  $\emph{I}$ is the laser intensity and
$\omega$ is the angular frequency) \cite{Keldysh2}. In the tunneling limit, where $\gamma$ approaches zero, the electron spectrum is the coherent superposition of the complex electron wave packet generated at each point in time of the laser pulse. In this quasistatic picture, the strength of each wavelet generated at time \emph{t} reaching the detector is the interference spectrum of the direct wave and the rescattered wave. In fact, based on classical trajectories and including the phase difference accumulated via the two paths, the ``side lobe" indeed can be qualitatively explained in terms of the interference pattern \cite{Bian2011,Huismans7}.  A theory based on quantum mechanics, however, turns out to be more difficult, especially by computing the side lobes as the hologram resulting from the interference of a reference wave and a signal wave. Numerical solution of the time-dependent
Schr\"{o}dinger equation (TDSE) is able to quantitatively reproduce
the experimental observation
\cite{Huismans7,Hickstein8,Meckel9,Marchenko11,Yang14,Bian2011}, but it is unable to explore the underlying mechanism of the side lobes. In Ref. [12], a quantal Coulomb-corrected SFA theory (CCSFA) has been used.  This is a semiclassical theory where the effect of Coulomb interaction between the target core and the electron has been approximately accounted for. The CCSFA results, however, do not compare very well with the experimental data or with the results from solving the TDSE.

  To understand the side lobes, here we simulated photoelectron spectra using the quantum-trajectory Monte Carlo (QTMC) method. QTMC was extended
from the classical-trajectory Monte Carlo method
\cite{Brabec1996PRA,Hu14,Chen2000} by including quantum
interference effect after the tunneling \cite{MinLi15}. The QTMC method has been successfully used to interpret photoelectron spectra in recent years with great success.
However, the QTMC method, which treats ionization rate under the quasistatic approximation, is only rigorously valid
in the limit of ${\gamma}\ll{1}$ \cite{Keldysh2}.

It has always been of great interest \cite{Eckle2,Boge5,Gkortsas6,IAIvanov14,CLWang14} to ask how important the nonadiabatic effect is in ionization since most experiments are carried out in the transition regime, or the so-called nonadiabatic tunneling regime
\cite{Yudin3,Yu4} where (${\gamma}\sim{1}$). Because side lobes have been observed not only in the tunneling regime
\cite{Huismans7} but also in the nonadiabatic tunneling regime
\cite{Huismans10,Marchenko11}, in this work we extend QTMC to account for nonadiabatic ionization, to study the side lobes. This generalized QTMC  is called GQTMC. We have found that GQTMC simulations are in much better agreement with experiment and with TDSE results. On the other hand, QTMC results are very close to CCSFA, and both do not reproduce either TDSE  or  experimental results.
 We further compare TDSE, QTMC and GQTMC results for side lobes in the multiphoton ionization and tunneling ionization regimes to establish that nonadiabatic effect of subcycle ionization dynamics are essential to the quantitative discription of photoelectron holography, but Coulomb effect is significant only in the multiphoton ionization regime.

\section{Methods}
 In the QTMC method, ionization is based on quantum subcycle adiabatic tunneling ionization theory of  Ammosov, Delone and Krainov \cite{Ammosov},  classical dynamics with combined laser and Coulomb fields \cite{Brabec1996PRA,Hu14,Chen2000}, and the Feynman's path integral approach \cite{MinLi15,Salieres2001}. The ionization rate is governed by
\begin{equation}
\Gamma_{qs}(t)=N_{qs}(t)\exp(-\frac{2(2I_p)^{3/2}}{3E_{0}f(t)|\cos{\Phi}(t)|}),
\end{equation}
where
\begin{equation}
N_{qs}(t)=A_{n^{*},l^{*}}B_{l,|m|}I_{p}(\frac{2(2I_{p})^{3/2}}{E_{0}f(t)|\cos{\Phi}(t)|})^{2n^{*}-|m|-1},
\end{equation}

In the GQTMC method, this expression is replaced by
\begin{equation}
\Gamma(t)=N(t)\exp(-\frac{E_{0}^{{2}}f^{{2}}(t)}{\omega^{{3}}}\Phi(\gamma(t),\theta(t))),
\end{equation}
where
\begin{equation}
N(t)=A_{n^{*},l^{*}}B_{l,|m|}(\frac{3\kappa}{\gamma^{3}})^{\frac{1}{2}}CI_{p}(\frac{2(2I_{p})^{3/2}}{E(t)})^{2n^{*}-|m|-1},
\end{equation}

The detailed meaning of the parameters in these equations are given in the supplementary information. The equations are given to show that the main difference between QTMC and GQTMC is mostly in the exponential factors in Eqs.(1) and (3). In QTMC, it is based on the quasistatic approximation so the exponential factor depends on the instantaneous time, while in GQTMC, the exponential factor at a given time is more complex since it accounts for the nonadiabatic ionization, as explained in the supplementary information. In the tunneling limit, $\gamma <<1$, the GQTMC is reduced to QTMC.

\begin{figure}[H]
\centering
\includegraphics[width=0.47\textwidth,height=0.2\textwidth,angle=0]{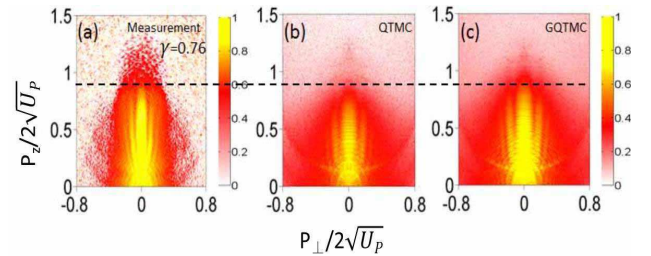}
\caption{Comparison of experimental two-dimensional photoelectron momentum distributions with calculations, for the metastable 6s state ($I_p=0.14$ a.u.) of xeon atom by lasers of wavelength of 7000 nm. (a): Experiment from Ref. \cite{Huismans7}  at $I=7.1\times{10^{11}}$ W/cm$^{2}$; (b) QTMC simulation, the laser pulse envelope is half-trapezoidal, constant for the first four cycles and ramped off linearly within the last two cycles, and the peak intensity is  $I=9.1\times{10^{11}}$ W/cm$^{2}$; (c): same as (b) but for GQTMC. The horizontal dashed line is the cutoff energy of the side lobe. The simulations included laser intensity distributions in the focused volume.}
\label{fig:false-color}
\end{figure}
\begin{figure*}[htb]
\centering
\includegraphics[width=0.7\textwidth,height=0.4\textwidth,angle=0]{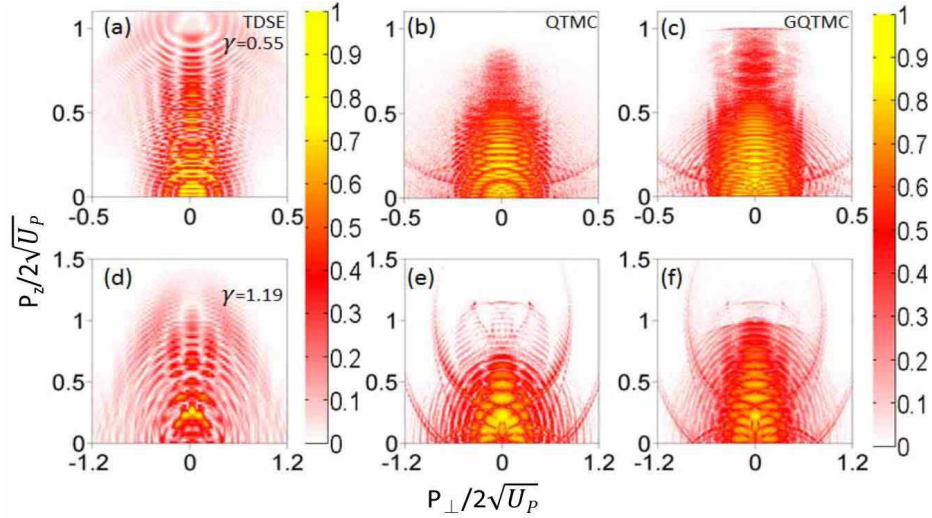}
\caption{Comparison of two-dimensional
photoelectron momentum spectra of Xenon atom from the ground state using TDSE, QTMC and GQTMC. Upper row:
 ${\gamma}=0.55$, $I=7.0\times{10^{13}}$ W/cm$^{2}$. The momentum in the vertical scale is normalized with respect to the cutoff momentum $2\sqrt{\textit{U}_{p}}$. Lower row: ${\gamma}=1.19$, $I=1.5\times{10^{13}}$ W/cm$^{2}$. The laser pulse envelope is half-trapezoidal, constant for the first six cycles and ramped off linearly within the
last four cycles. Laser wavelength: $\lambda=1700$ nm. The results show that side lobes seen in the TDSE and in GQTMC are less prominant in the QTMC, and the cutoff energy of the side lobe is also smaller in QTMC than from TDSE and GQTMC, illustrating that nonadiabatic ionization effect is nonnegligible even for  ${\gamma}=0.55$.}
\label{fig:false-color}
\end{figure*}

After ionization, the classical motion of the electron in the combined laser and Coulomb fields is governed by Newton equation:
\begin{equation}
\frac{d^{2}}{dt^{2}}\mathbf{r}=-\mathbf{E}(t)-\bigtriangledown(\mathrm{V}(\mathbf{r})).
\end{equation}
Here, V(\textbf{r})  is the potential of the ion. Each electron trajectory is weighted by the ionization rate $\Gamma(t_0,v_{\perp0})=\Gamma(t_0)\times\Omega(v_{\perp0})$ in which
$\Omega(v_{\perp0})\propto[\sqrt{2I_{p}}/|E(t_{0})|]\exp[\sqrt{2I_{p}}(v_{\perp0})^{2}/|E(t_{0})|]$. Note that ``$\perp$" is the direction perpendicular to the laser polarization axis.

In addition to modification of the ionization rate, we consistently substitute the coordinate of the tunnel exit point by $z_{0}=\frac{2I_{p}}{E(t_{0})}(1+\sqrt{1+\gamma^{{2}}(t_{0})})^{-{1}}$ ($I_{p}$ is the ionization
potential, $E(t_{0})$ is the initial instantaneous field, and $\gamma{(t_{0})}$ is the Keldysh parameter depending on the instantaneous time.) which shifts the exit point toward the atomic core due to the nonadiabatic effect \cite{Perelomov}. According to  Feynman's path integral approach, the $\emph{j}$th electron acquires a phase given by the classical action along the trajectory: \cite{MinLi15,Salieres2001}
\begin{equation}
{S_j}(\mathbf{p},t_0 )=\int_{t_0}^{+\infty}\{\mathbf{v_p}^{2}(\tau)/2+I_p-1/|\mathbf{r}(t)| \}d\tau,
\end{equation}
where $\textbf{p}$ is the asymptotic momentum of the $\emph{j}$th electron. The probability of each asymptotic momentum is determined by
\begin{equation}
{|\Psi|_\mathbf{p}}^{2}=|\sum_j\sqrt{\Gamma(t_0,v_{r}^j ) }\exp(-iS_j(\mathbf{p},t_{0}))|^{{2}}.
\end{equation}
Using a parallel algorithm, the PMD was obtained with one billion electron trajectories.

\section{Results}
\subsection{ Comparison of simulations with experiment}

 Fig. 1 shows the experimental PMD obtained from Huismans et al \cite{Huismans7} where PMD was reported for ionization of the metastable 6s electron of Xe by a 7000 nm laser with peak intensity of  $I=7.1\times{10^{11}}$ W/cm$^{2}$, corresponding to ${\gamma}=0.76$. In the QTMC and GQTMC calculations, like in experiment, we have integrated electron signals from the whole focal volume. To our knowledge, this is the first time that the data of Huismans \emph{et al.} are compared to simulations that includes volume integration. We found that the experimental data best agree with GQTMC if the peak laser intensity is about $9.1\times{10^{11}}$ W/cm$^{2}$, about 30 percent higher than the value cited in the experiment. Note that the cutoff energy in the GQTMC simulation is in good agreement with the experimental data, while simulation with the same laser intensities using QTMC always has a lower energy cutoff. The side lobes are clearly seen in all three frames. Besides the main lobe along the polarization axis, there are two lobes on each side of the axis, with the outer 2nd lobe is weaker. The simulations show one strong arc on each side nearly perpendicular to the polarization axis which are not clearly seen in the experimental data.

\subsection{Nonadiabatic effects}

Fig. 2 shows that nonadiabatic ionization is important for describing side lobes correctly. Two laser intensities are used, one for ${\gamma}=0.55$ and another for
${\gamma}=1.19$. Xeon atom in the ground state was used  and QTMC and GQTMC are used together with solutions from TDSE \cite{Yang12,Yang14}. One can see that side lobes are much clearly seen in TDSE and GQTMC than in QTMC. The cutoff energy in QTMC also tends to be lower than the other two. These single intensity results demonstrate that accurate description of side lobes needs to account for the nonadiabatic ionization effect. This is qualitatively understood by the fact that the side lobes include contribution from higher energy direct electrons. In the static theory, they would be ionized when the field is weak and the vector potential is larger. The nonadiabatic effect enhances ionization yields, especially when the laser field is near zero. Thus nonadiabatic ionization effect is enhanced for the formation of side lobes. In Fig. 2, these observations hold for both the tunneling regime and the multiphoton regime.

\begin{figure}[H]
\centering
\includegraphics[width=0.45\textwidth,height=0.45\textwidth,angle=0]{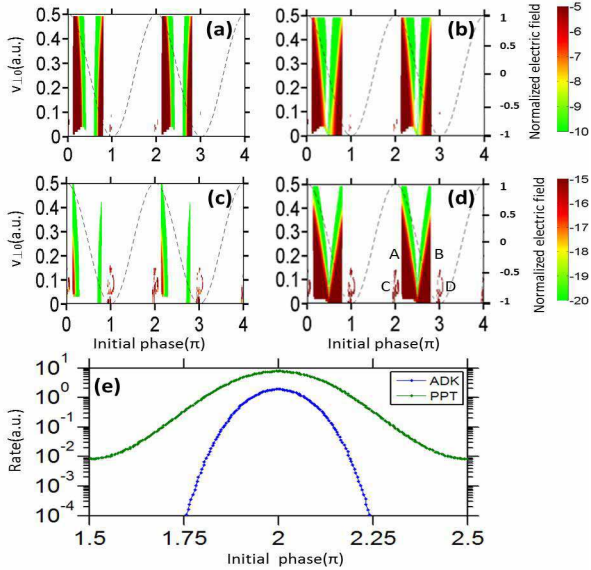}
\caption{Distributions of the initial transverse
velocities and the initial ionization phases of the laser. Left column: QTMC
simulations; right column: GQTMC simulations. The dash line is the
laser electric field. Upper row:  ${\gamma}=0.55$. Lower row : ${\gamma}=1.19$. The simulation parameters are the same as in Fig. 2.
(e) comparison of total ionization rate using ADK theory and the PPT model for ${\gamma}=1.19$, vs the phase of the laser field.
Difference due to nonadiabatic ionization is much larger when the laser electric field is small.}
\label{fig:false-color}
\end{figure}

To further explore the nonadiabatic ionization effect we performed  quantum trajectories analysis
of the PMD for both ${\gamma}=0.55$ and ${\gamma}=1.19$,  as shown in Fig. 3. We analyze the normalized momentum interval of
${\textit{P}_{z}}$$\geq{{\frac{1}{2}}\textit{P}_{z_{cutoff}}}$,
with ${\textit{P}_{z_{cutoff}}}$=$2\sqrt{\textit{U}_{p}}$ being
the momentum corresponding to the $2\textit{U}_{p}$ cutoff in
energy. In this region, the momentum distributions are dominated
by the side lobes structure. The QTMC and GQTMC models offer the opportunity to trace back the initial transverse velocity
and the initial laser phase for each electron contributing to a given
momentum distribution spot. By comparing these two models, the influence of nonadiabatic ionization can be revealed.
In Fig. 3(d), over half an optical period (laser phase from 2$\pi$ to 3$\pi$), we used A, B, C, D to denote the laser phase of ionization vs the distribution of the initial transverse velocity. The electrons generated at each of such a half optical cycle will interfere. Similar plots are also shown in (a) to (c). The differences among these frames are due to the degree of nonadiabaticity in tunnel ionization.

 For ${\gamma}=0.55$, the initial phase distributions are very similar
for the QTMC and GQTMC models, but the areas A and B
calculated by the GQTMC method are slightly wider than those by
the QTMC method. For the case of ${\gamma}=1.19$, dramatic change occurs
 in these two areas. The much broader width in GQTMC as compared to QTMC  reflects the subcycle ionization dynamics in nonadiabatic ionization, where ionization does not follow the instantaneous electric field of the laser pulse, but ``spreads" out over a broader time interval, thus resulting in a much sharper contrast in subcycle ionization dynamics in the multiphoton ionization regime, as clearly seen between (c) and (d) where ${\gamma}=1.19$. Fig. 3(e) compares the subcycle ionization rates over one quarter cycle according to ADK \cite{Yudin3,Ammosov}model vs the PPT model
\cite{Perelomov}. Note in Fig. 3(e), the ionization includes all energies of the photoelectrons, while Figs. (3a) to 3(d) include only higher electron energies described in the previous paragraph.

\begin{figure}[H]
\centering
\includegraphics[width=0.45\textwidth,height=0.35\textwidth,angle=0]{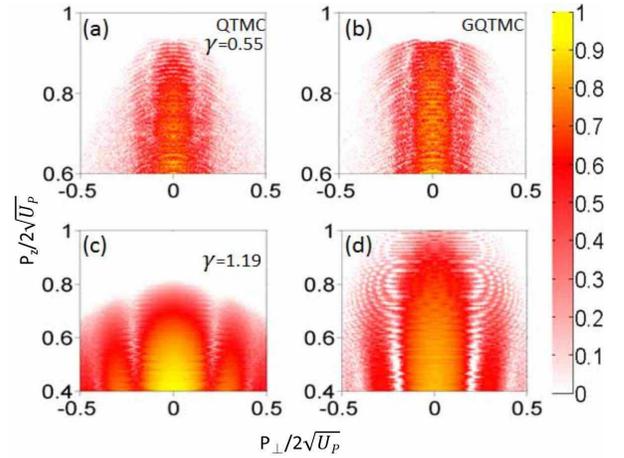}
\caption{Comparison of the final PMD due to ionization from area A (see Fig. 3d) calculated by the QTMC and GQTMC for ${\gamma}=0.55$ and $1.19$, respectively. The parameters are the
same as in Fig. 2.}
\label{fig:false-color}
\end{figure}

\begin{figure*}[htb]
\centering
\includegraphics[width=0.7\textwidth,height=0.4\textwidth,angle=0]{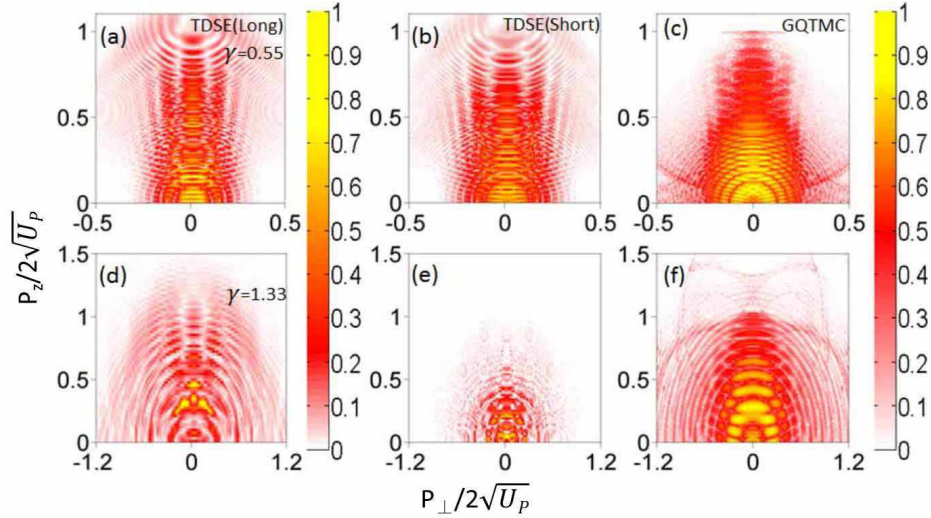}
\caption{ Effect of long range Coulomb potential on the side lobes. Upper row for $\gamma=0.55$, $I=5.0\times{10^{13}}$ W/cm$^{2}$, $\lambda=2000$ nm. Lower row for $\gamma=1.33$, $I=1.2\times{10^{13}}$
W/cm$^{2}$, $\lambda=1700$ nm. Left column: with long-range
Coulomb potential. Middle and right column: with short-range
potential. Side lobes seen with or without Coulomb potential for $\gamma=0.55$.  Weak side lobes can be seen
in TDSE with Coulomb potential, but not for short range potential for both TDSE and GQTMC, for $\gamma=1.33$.
In the multiphoton ionization regime, Coulomb potential is needed to incur large angle collision for the direct process for the side lobe to appear. }
\label{fig:false-color}
\end{figure*}

It has been demonstrated that side lobes are resulting from the interference between direct and rescattered electron wave packets that are emitted during the same quarter cycle of the laser field\cite{Huismans7}, i.e., from  area A in Fig. 3(d) . To demonstrate this, we show in Fig. 4 the calculated PMD
for electrons emitted from area A in each of the four cases in Fig. 3. Consistent with the previous SFA and classical calculations \cite{Huismans7, Bian2011}, electrons emitted from area A indeed yield the side lobes as holographic interference patterns (see Fig. 4). For ${\gamma}=0.55$, both methods reproduce well the
side lobes seen from  the TDSE calculation. For ${\gamma}=1.19$, the QTMC method underestimates the cutoff energy of the side lobes.
Further examination of the initial velocity distributions of electrons emitted from area A for ${\gamma}=1.19$ (see Figs. 3 (c) and (d)), one finds that the mismatch between QTMC and GQTMC is much more severe when the instantaneous electric field is around zero. The quasi-static ADK theory underestimates the ionization yield in this part of the laser field when ${\gamma}\sim1$.  Due to nonadiabatic tunneling, substantial ionization can occur even when the instantaneous electric field is weak\cite{Yudin3, Yu4}. Since subcycle ionization dynamics is well described by Eq. (4) \cite{Yudin3} which is the basis of our GQTMC model,  the PMD's calculated via GQTMC are able to describe accurately the side lobes observed in the experiments or from the TDSE calculations over a large range of ${\gamma}$.

  Next consider ionization from other laser phase areas B, C, and D.   For example, ionization from C and D occurs near the field crest, they show no evidence of nonadiabatic tunneling ionization, and QTMC and GQTMC results are essentially identical, see Fig. 3.  Ionization from C and D becomes significantly more important for ${\gamma}=1.19$  in the multiphoton ionization regime. Furthermore, it has been demonstrated that electron trajectories in area C correspond to a form of transverse backward-scattering driven by the Coulomb field, which induces a totally different interference structure from that of the PH \cite{Song}. In the QTMC simulation, since the cutoff energy induced by area A is low, interference fringes induced by electrons from area C can be seen beyond the side lobes, see Fig. 2(e). However for GQTMC simulation, these interference fringes are only observable at large transverse final momentum where the side lobes are much weaker (see Fig. 2(f)). It should be mentioned here that, similar to area B, electrons from area D, which come from the negative electric field, only contribute to the background of the fringes.

\subsection{Long-range Coulomb potential effects}

In the following, we explore the role of Coulomb potential in the holographic
interference structures. We calculate PMDs by the TDSE and GQTMC
with the long-range Coulomb potential and a short-range potential
\cite{shortrange} for different laser intensities and wavelengths shown in Fig. 5.
For $\gamma=0.55$, the side lobes can be clearly seen from TDSE with Coulomb or with short-range potentials. In GQTMC, the
side lobes are still distinguishable with short-range potential (Fig. 5(c)), but they are more pronounced with Coulomb potential (Fig. 2(f)).
In contrast, for $\gamma=1.33$, the interference pattern can be seen from TDSE with Coulomb potential, but with short-range potential, side lobes are hardly discernable in both   TDSE and  GQTMC simulations. These results demonstrate that Coulomb potential plays a significant role for observing side lobes or holographic interference at lower laser intensity, but not at higher laser intensity.

To explain these observations, we show statistics of the
longitudinal momentum of the rescattering electrons, for both long-range and
short-range potentials, see Fig. 6. For $\gamma=0.55$, the
 long-range potential tends to populate electrons with larger longitudinal momentum (Fig. 6(a)), but the difference is not large. Moreover,
the probability of scattering with large change of transverse
momentum, which is essential for the formation of the interference
fringe, is also reduced in the short-range potential (see typical trajectories shown in Fig.
6(c)). Therefore, the fringes become less apparent in the
short-range potential than that in the long-range potential.
More interestingly, in the nonadiabatic tunneling regime
($\gamma=1.33$), Fig. 6(b) shows that Coulomb potential can increase the distribution for large
longitudinal momentum for the rescattering electron  nearly by a factor of two beyond $P_z\sim
0.5$ in this case. For weaker laser field and/or shorter wavelength, the quiver amplitude ($\alpha=E_0/\omega^2$) of the electron is much smaller, thus the electron is more prone to be pulled back to collide with the core in the presence of long-range
potential.  Fig. 6(d) shows two sets of trajectories of electrons where the Coulomb potential significantly modifies the motion as compared to short range potential, thus explaining the significant drop in electrons contributing to the formation of side lobes, as shown in Fig. 6(b).

\begin{figure}[H]
\centering
\includegraphics[width=0.45\textwidth,height=0.35\textwidth,angle=0]{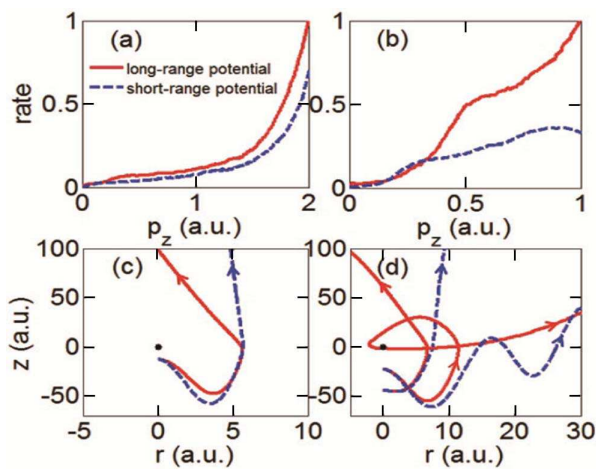}
\caption{(Color online) Effect of Coulomb potential vs short-range potential on the final longitudinal momentum distribution of
the rescattering  electrons for  $\gamma=0.55$ (left column) and 1.33 (right column), respectively. Typical electron trajectories are shown for these two regimes also.  Note that z is the longitudinal direction and r is the transverse direction. The parameters are the same as in Fig. 5.}
\label{fig:false-color}
\end{figure}

\section{Discussion}
  In summary, a GQTMC model has been used to calculate the two-dimensional photoelectron momentum distributions.
The model was used to describe accurately the so-called side lobes which had been understood as photoelectron hologram. Like optical hologram, it is due to the interference between a reference wave and a signal wave. The hologram offers the opportunity to probe the structure of the object. In side lobes, the reference wave is the direct electron emission while the signal wave is due to electrons that have been rescattered by the ion. Unlike optical hologram, however, in photoelectron hologram, both the reference and signal waves are influenced by the subcycle ionization mechanism and by the subsequent interaction due to the laser field and the Coulomb potential. To use photoelectron holography to probe the target structure, both of these effects have to be examined first. This is accomplished in the present work using the GQTMC method. In this method, the direct wave and the signal wave can be separated. Since the reference wave is nearly independent of the target, one expects to be able to retrieve target structure from the signal wave. The present work will make such retrieval possible, but with photoelectrons that suffer harder collisions, i.e., at angles beyond the side lobes.  After this extension has been made, photoelectron holography may offer as a complementary tool for imaging the dynamics of structural changes at femtosecond timescale, similar to laser induced electron diffraction (LIED) \cite{Xu2010, Blaga2012, Pullen2015}. The LIED method used backscatted photoelectrons in the high-energy region where there are no contributions from direct wave. The method is easier to analyze but it suffers from low yields. Photoelectron holography method uses lower energy electrons where the ionization yield is much larger. Using GQTMC method, the direct wave can be extracted from the full spectrum. Further analysis of the signal wave would allow the determination of target structure. To make such a method possible, clearly photoelectron holography should first be investiaged on molecular targets. It is also worth mentioning that photoelectron holography by long wavelength lasers discussed here is similarly used in ionization of molecules by hard X-rays. In this case, a hard X-ray is used to ionize the inner shell of a specific atom. The photoelectron generated may reach the detector directly, or via scattering with neighboring atoms. The interference of these two waves offer the opportunity to extract the target moleucles. This method is being explored with X-ray free-electron lasers (XFEL's) recently\cite{Krasniqi2010}.

\section*{Funding Information}
The work was supported by the National Basic Research Program of
China Grant (No. 2013CB922201), the NNSF of
China (Grant Nos. 11374202, 11274220, 11274219, 11274050, 11334009 and
11425414), Guangdong NSF (Grant No. 2014A030311019), and the Open Fund of the
State Key Laboratory of High Field Laser Physics (SIOM). W. Y. is supported by the ``YangFan" Talent Project of Guangdong Province. C. D. Lin is partially supported by Chemical Sciences, Geosciences and Biosciences Division, Office of Basic Energy Sciences, Office of Science, U. S. Department of Energy under Grant No. DE-FG02-86ER13491.

\section*{Acknowledgments}

We are grateful to M. Vrakking and Y. Huismans for providing us
their data of Ref\ . \cite{Huismans7} in a numeric format and for fruitful discussions. We also thank X. Liu for illuminating discussions.

\section*{Author contributions statement}
C.L., P.L. and Z.S. performed all the calculations with help and discussions with X.S., Z.C., W.Y., S.H., C.D.L., and J.C.; X.S., W.Y., C.D.L. and J.C. wrote the
manuscript. All authors contributed to finalizing and approving the manuscript.

\section*{Additional information}
Competing financial interests: The authors declare no competing financial interests.





\begin{thebibliography}{xx}


\bibitem{ap79} P. Agostini, F. Fabre, G. Mainfray, G. Petite, and N. K. Rahman, ``Free-Free Transitions Following Six-Photon Ionization of Xenon Atoms,'' Phys. Rev.
Lett. \textbf{42}, 1127--1130 (1979).

\bibitem{wbecker1} W. Becker, F. Grasbon, R. Kopold, D. B. Milo\v{s}evi\'c, G. G. Paulus, and H. Walther, ``Above-Threshold Ionization: From Classical Features to Quantum Effects,'' Adv. At. Mol. Opt. Phys. \textbf{48}, 35--98 (2002).



\bibitem{Corkum1} P. B. Corkum, ``Plasma perspective on strong field multiphoton ionization,'' Phys. Rev. Lett. \textbf{71}, 1994--1997 (1993).

\bibitem{Blaga2009NatPhys} C. I. Blaga, F. Catoire, P. Colosimo, G. G. Paulus, H. G. Muller, P. Agostini, and L. F. DiMauro, ``Strong-field photoionization revisited,'' Nature Phys. \textbf{5}, 335--338 (2009)

\bibitem{Quan2009PRL} W. Quan, Z. Lin, M. Wu, H. Kang, H. Liu, X. Liu, J. Chen, J. Liu, X. He, S. Chen, H. Xiong, L. Guo, H. Xu, Y. Fu, Y. Cheng, and Z. Xu, ``Classical Aspects in Above-Threshold Ionization with a Midinfrared Strong Laser Field,'' Phys. Rev. Lett.  \textbf{103}, 093001 (2009).

\bibitem {Wu2012PRL} C. Wu, Y. Yang, Y. Liu, Q. Gong, M. Wu, X. Liu, X. Hao, W. Li, X. He, and J. Chen, ``Characteristic Spectrum of Very Low-Energy Photoelectron from Above-Threshold Ionization in the Tunneling Regime,'' Phys. Rev. Lett. \textbf{109}, 043001 (2012).
\bibitem{liu10} C. Liu and K. Z. Hatsagortsyan, ``Origin of Unexpected Low Energy Structure in Photoelectron Spectra Induced by Midinfrared Strong Laser Fields,'' Phys. Rev. Lett. \textbf{105}, 113003 (2010).

\bibitem{yan10} T. Yan, S. V. Popruzhenko, M. J. J. Vrakking, and D. Bauer, ``Low-Energy Structures in Strong Field Ionization Revealed by Quantum Orbits,'' Phys. Rev. Lett. \textbf{105}, 253002 (2010).

\bibitem{rost12} A. K\"astner, U. Saalmann, and J. M. Rost, ``Electron-Energy Bunching in Laser-Driven Soft Recollisions,'' Phys. Rev. Lett. \textbf{108}, 033201 (2012).

\bibitem{chu2012PRA} W. Chu, M. Wu, B. Zeng, J. Yao, J. Ni, H. Xiong, H. Xu, Z. Lin, H. Kang, W. Quan, J. Chen, X. Liu, Y. Cheng, and Z. Xu, ``Unexpected breakdown of the simple man's model for strong-field photoionization in the high-energy recollision region,'' Phys.  Rev.  A \textbf{85}, 021403(R) (2012).

\bibitem{Guo2013} L. Guo, S. Han, X. Liu, Y. Cheng, Z. Xu, J. Fan, J. Chen, S. G. Chen, W. Becker, C. I. Blaga, A. D. DiChiara, E. Sistrunk, P. Agostini, and L. F. DiMauro,
    ``Scaling of the Low-Energy Structure in Above-Threshold Ionization in the Tunneling Regime: Theory and Experiment,'' Phys. Rev. Lett. \textbf{110}, 013001 (2013).
\bibitem{Huismans7} Y. Huismans, A. Rouze, A. Gijsbertsen, J. H. Jungmann, A. S. Smolkowska, P. S. W. M. Logman, F. L\'epine, C. Cauchy, S. Zamith, T. Marchenko, J. M. Bakker, G. Berden, B. Redlich, A. F. G. van der Meer, H. G. Muller, W. Vermin, K. J. Schafer, M. Spanner, M. Yu. Ivanov, O. Smirnova, D. Bauer, S. V. Popruzhenko, and M. J. J. Vrakking, ``Time-Resolved Holography with Photoelectrons,'' Science \textbf{331}, 61--64 (2011).

\bibitem{Hickstein8} D. D. Hickstein, P. Ranitovic, S. Witte, X. Tong, Y.  Huismans, P. Arpin, X. Zhou, K. E. Keister, C. W. Hogle, B. Zhang, C. Ding, P. Johnsson, N. Toshima, M. J. J. Vrakking, M. M. Murnane, and H. C. Kapteyn,  ``Direct Visualization of Laser-Driven Electron Multiple Scattering and Tunneling Distance in Strong-Field Ionization,'' Phys. Rev. Lett. \textbf{109}, 073004 (2012).

\bibitem{Meckel9} M. Meckel, A. Staudte, S. Patchkovskii, D. M. Villeneuve, P. B. Corkum, R. D\"orner, and M. Spanner, ``Signatures of the continuum electron phase in molecular strong-field photoelectron holography,'' Nature Phys. \textbf{10}, 594--600 (2014).

\bibitem{Huismans10} Y. Huismans, A. Gijsbertsen, A. S. Smolkowska, J. H. Jungmann, A. Rouz\'ee, P. S. W. M. Logman, F. L\'epine, C. Cauchy, S. Zamith, T. Marchenko, J. M. Bakker, G. Berden, B. Redlich, A. F. G. van der Meer, M. Yu. Ivanov, T. Yan, D. Bauer, O. Smirnova, and M. J. J. Vrakking, ``Scaling Laws for Photoelectron Holography in the Midinfrared Wavelength Regime,'' Phys. Rev. Lett. \textbf{109}, 013002 (2012).

\bibitem{Marchenko11} T. Marchenko, Y. Huismans, K. J. Schafer, and M. J. J. Vrakking, ``Criteria for the observation of strong-field photoelectron holography,'' Phys. Rev. A \textbf{84}, 053427 (2011).

\bibitem{Yang12} W. Yang, X. Song, and Z. Chen, ``Phase-dependent above-barrier ionization of excited-state electrons,'' Opt. Express \textbf{20}, 12067--12705 (2012).

\bibitem{Yang14} W. Yang, Z. Sheng, X. Feng, M. Wu, Z. Chen, and X. Song, ``Molecular photoelectron holography with circularly polarized laser pulses,'' Opt. Express \textbf{22}, 2519--2527 (2014).

\bibitem{Bian2011} X. Bian, Y. Huismans, O. Smirnova, K. Yuan, M. J. J. Vrakking, and A. D. Bandrauk, ``Subcycle interference dynamics of time-resolved photoelectron holography with midinfrared laser pulses,'' Phys. Rev. A \textbf{84}, 043420 (2011).

\bibitem{Keldysh2} L. V. Keldysh, ``Ionization in the Field of a Strong Electromagnetic Wave,'' Sov. Phys. JETP. \textbf{20}, 1307--1314 (1965).
\bibitem{Brabec1996PRA} T. Brabec, M. Yu. Ivanov, and P. B. Corkum, ``Coulomb focusing in intense field atomic processes,'' Phys. Rev. A \textbf{54}, R2551--R2554 (1996).

\bibitem{Hu14} B. Hu, J. Liu, and S. Chen, ``Plateau in above-threshold-ionization spectra and chaotic behavior in rescattering processes,'' Phys. Lett. A \textbf{236}, 533--542 (1997).

\bibitem{Chen2000} J. Chen, J. Liu, and S. Chen, ``Rescattering effect on phase-dependent ionization of atoms in two-color intense fields,'' Phys. Rev. A \textbf{61}, 033402 (2000).

\bibitem{MinLi15} M. Li, J. Geng, H Liu, Y. Deng, C. Wu, L. Peng, Q. Gong, and Y. Liu, ``Classical-Quantum Correspondence for Above-Threshold Ionization,'' Phys. Rev. Lett. \textbf{112}, 113002 (2014).

\bibitem{Eckle2}P. Eckle, A. N. Pfeiffer, C. Cirelli, A. Staudte, R. D\"orner, H. G. Muller, M. B\"uttiker, and U. Keller, ``Attosecond Ionization and Tunneling Delay Time Measurements in Helium,'' Science \textbf{322}, 1525--1529 (2008).

\bibitem{Boge5} R. Boge,  C. Cirelli, A. S. Landsman, S. Heuser, A. Ludwig, J. Maurer, M. Weger, L. Gallmann, and U. Keller, ``Probing Nonadiabatic Effects in Strong-Field Tunnel Ionization,'' Phys. Rev. Lett. \textbf{111}, 103003 (2013).

\bibitem{Gkortsas6} V. M. Gkortsas, S. Bhardwaj, C. Lai, K. H. Hong, E. L. F. Filho, and F. X. K\"artner, ``Interplay of mulitphoton and tunneling ionization in short-wavelength-driven high-order harmonic generation,'' Phys.  Rev.  A \textbf{84}, 013427 (2011).

\bibitem{IAIvanov14} I. A. Ivanov and A. S. Kheifets, ``Strong-field ionization of He by elliptically polarized light in attoclock configuration,'' Phys.  Rev.  A \textbf{89}, 021402(R) (2014).

\bibitem{CLWang14} C. Wang, X. Lai, Z. Hu, Y. Chen, W. Quan, H. Kang, C. Gong, and X. Liu, ``Strong-field atomic ionization in elliptically polarized laser fields,'' Phys.  Rev.  A \textbf{90}, 013422 (2014).

\bibitem{Yudin3} G. L. Yudin and M. Yu. Ivanov, ``Nonadiabatic tunnel ionization: Looking inside a laser cycle,'' Phys. Rev. A \textbf{64}, 013409 (2001).

\bibitem{Yu4} M. Y. Ivanov, M. Spanner, and O. Smirnova, ``Anatomy of strong field ionization,'' J. Mod. Opt. \textbf{52}, 165--184 (2005).

\bibitem{Ammosov} M. V Ammosov, N. B. Delone, and V. P. Krainov, ``Tunnel ionization of complex atoms and of atomic ions in an alternating electromagnetic field,'' Sov. Phys. JETP \textbf{64}, 1191--1194 (1986).


\bibitem{Salieres2001} P. Sali\`{e}res, B. Carr\`{e}, L. Le D\`{e}roff, F. Grasbon, G. G. Paulus. H. Walther, R. Kopold, W. Becker, D. B. Milo\v{s}evi\'c, A. Sanpera, and M. Lewenstein, ``Feynman's Path-Integral Approach for Intense-Laser-Atom Interactions,'' Science \textbf{292}, 902--905 (2001).
\bibitem{Perelomov} A. M. Perelomov, V. S. Popov, and M. V. Terent'ev, ``Ionization of Atoms in an Alternating Electric Field,'' Zh. $\acute{E}$ksp. Teor. Fiz. \textbf{50}, 1393 (1966) [Sov. Phys. JETP \textbf{23}, 924--934 (1966)].

\bibitem{Song} X. Song, P. Liu, C. Lin, Z. Sheng, X. Yu, W. Yang, S. Hu, J. Chen, S. Xu, Y. Chen, W. Quan, and X. Liu, ``Attosecond Interference Induced by Coulomb-Field-Driven Transverse Backward-Scattering Electron Wave-Packets,'' submitted (2015).

\bibitem{shortrange} Z. Chen, T. Morishita, A. T. Le, M. Wickenhauser, X. M. Tong, and C. D. Lin, ``Analysis of two-dimensional photoelectron momentum spectra and the effect of the long-range Coulomb potential in single ionization of atoms by intense lasers,'' Phys. Rev. A \textbf{74}, 053405 (2006).
\bibitem{Xu2010} J. Xu, Z. Chen, A. T. Le,and C. D. Lin, ``Self-imaging of molecules from diffraction spectra by laser-induced rescattering electrons,'' Phys. Rev. A \textbf{82},033403 (2010).


\bibitem{Blaga2012} C. I. Blaga,et al. ``Imaging ultrafast molecular dynamics with laser-induced electron difftraction,'' Nature \textbf{483},194–197(2012).

\bibitem{Pullen2015} M. G. Pullen, B. Wolter, A. T. Le, M. Baudisch, M. Hemmer,A. Senftleben, C. D. Schr\"oter, J. Ullrich, R. Moshammer, C. D. Lin et al.,``Imaging an aligned polyatomic molecule
    with laser-induced electron diffraction,'' Nat Commun \textbf{6}, 7262 (2015).

\bibitem{Krasniqi2010} F. Krasniqi,et al.``Imaging molecules from within: ultrafast angstr\"om-scale
    structure determination of molecules via photoelectron holography using freeelectron lasers,'' Phys. Rev. A \textbf{81},033411 (2010).


\end{thebibliography}


\end{document}